\documentclass[useAMS,usenatbib,referee]{mn2e}
\usepackage{graphicx,graphics,aas_macros,amsmath}
\usepackage[authoryear]{natbib}
\title[4U 0114+65]{Is 4U 0114+65 an eclipsing HMXB?}
\author[P. Pradhan, B.Paul, B.C. Paul, E. Bozzo, T. Belloni]{Pragati Pradhan$^{1,2}$ \thanks{E-mail:pragati2707@gmail.com;}, Biswajit Paul$^{3}$, B.C. Paul$^{2}$, Enrico Bozzo$^{4}$, Tomaso. M. Belloni$^{5}$ \\
$^{1}$ St. Joseph's College, Singamari, Darjeeling-734104, West Bengal, India\\
$^{2}$ North Bengal University, Raja Rammohanpur,  District Darjeeling-734013, West Bengal, India \\
$^{3}$ Raman Research Institute, Sadashivnagar, Bangalore-560080, India\\
$^{4}$ ISDC, University of Geneva, Chemin d'Ecogia 16, Versoix, 1290, Switzerland \\
$^{5}$ INAF Osservatorio Astronomico di Brera, Via E Bianchi 46, 23807 Merate (LC), Italy
 }
\begin{document}
\date{}
\maketitle
\label{firstpage}
\begin{abstract}
We present the pulsation and spectral characteristics of the HMXB 4U 0114+65 during a \emph{Suzaku} observation covering the part of the orbit 
that included the previously known low intensity emission of the source (dip) and the egress from this state. This dip has been interpreted in previous works 
as an X-ray eclipse. Notably, in this Suzaku observation, the count rate during and outside the dip vary by a factor of only 2-4 at odds with 
the eclipses of other HMXBs, where the intensity drops upto two orders of magnitude. The orbital intensity profile of 4U 0114+65 is 
characterized by a narrow dip in the RXTE-ASM (2-12 \rm{keV}) light curve and a shallower one in the Swift-BAT (15-50 \rm{keV}), which is different from eclipse 
ingress/egress behaviour of other HMXBs. The time-resolved spectral analysis reveal moderate absorption column density 
(N$_{H}$ - 2-20 $\times$ $10^{22}$ atoms $cm^{-2}$) and a relatively low equivalent width ($\sim$ 30 \rm{eV} \& 12 \rm{eV} of the iron K$_\alpha$ and K$_\beta$ 
lines respectively) as 
opposed to the typical X-ray spectra of HMXBs during eclipse where the equivalent width is $\sim$ 1 \rm{keV}. Both XIS and PIN data show clear pulsations during the dip, 
which we have further confirmed using the entire archival data of the IBIS/ISGRI and JEM-X instruments
onboard \emph{INTEGRAL}. The results we presented question the previous interpretation of the dip in the light curve of 4U 0114+65 as an 
X-ray eclipse. We thus discuss alternative interpretations of the periodic dip in the light curve of 4U 0114+65.
\end{abstract}
\section{Introduction}

\label{sect:intro}
\rm{}
The hard X-ray source 4U 0114+65 (alias 2S 0114+650) is a pulsar belonging to a class of persistent High-Mass X-ray Binaries (HMXBs), showing properties similar 
to both Be \citep{koenigsberger1983} and supergiant X-ray 
binaries \cite[]{crampton1985,reig1996}. It was discovered
in the SAS-3 galactic survey \citep{dower1977}. The pulsar's companion is a B1Ia supergiant (LS I+65 010) located at a distance of $\sim$ 7.2 \rm{kpc} \citep{reig1996}. From the optical radial velocity measurements,
\cite{crampton1985} first confirmed the binary nature of 4U 0114+65. 
By assuming an eccentricity of $\sim$0.16, these authors estimated an orbital period of $\sim$11.588 \rm{d}. 
The corresponding value in case of a circular orbit is $\sim$11.591 \rm{d}.
Similar values of the system orbital period have also been reported in later works:
 \cite{corbet1999} obtained an X-ray variability of $\sim$ 11.63 \rm{d} from the \emph{RXTE}-ASM observations carried out between 1996-1999; \cite{wen2006} 
 reanalysed ASM data from 1996-2004
and obtained an orbital period of $\sim$ 11.599 \rm{d}; \cite{grundstrom2007} obtained an orbital period of $\sim$
11.5983 \rm{d} by using both optical and X-ray observations. These authors also reported an eccentricity of the orbit of $\sim$ 0.18.  \\
A stable 2.78 \rm{h} periodicity in the X-ray light curve of the source was first reported by \cite{finley1992} by using \emph{EXOSAT} 
and \emph{ROSAT} data. 
A similar modulation was also observed in the optical by \cite{taylor1995}. 
\cite{corbet1999} and \cite{hall2000} further confirmed the same periodicity by analyzing the \emph{RXTE}-ASM and PCA data, respectively. 
Additional X-ray observations proved that the periodicity at 2.7 \rm{h} was getting faster over time, firmly establishing its association with the 
spin period of the neutron star hosted in this system \citep[]{hall2000,bonning2005,farrell2008}. 
This is one of the slowest pulsar known and various models have been proposed to explain its unusually long spin period (\citealt{hall2000} and references therein). \cite{li1999} suggested that such long spin period could be achieved if the neutron star was initially born as a magnetar. 
The possible cyclotron resonant absorption features at $\sim$ 22 and 44 \rm{keV} reported by \cite{bonning2005} did not provide support in favor of this hypothesis, as
they would imply a magnetic field of $\sim$ 2.5$\times$10$^{12}$ \rm{G} for the pulsar in 4U 0114+65. 
However, subsequent observations did not confirm the detection of the cyclotron features \citep{denhartog2006,masetti2006, farrell2008}.\\ 
4U 0114+65 shows a remarkable variability in X-rays, with aperiodic flares lasting for a few hours \cite[]{yamauchi1990,apparao1991} and short-term flickering
occuring on timescales of minutes \citep{koenigsberger1983}. A superorbital modulation of $\sim$ 30.7 \rm{d} has been detected in \emph{RXTE}-ASM data 
\citep{farrell2006, wen2006}, and found to be stable over time. This makes 4U 0114+65 the fourth system to show stable superorbital variability 
(see also the cases of SMC X-1, Her X-1, LMC X-4 and SS 433; \citealt{sood2006}).\\
The orbital intensity profile of 4U 0114+65 as obtained from the \emph{RXTE}-ASM shows a dip that has been usually associated with an X-ray eclipse.
Here we report on the timing and broad band spectral characteristics of the X-ray emission from 4U 0114+65 as observed with \emph{Suzaku} 
during part of the dip and the dip egress. The \emph{Suzaku} observation indicates that the X-ray emission from the source recorded during the dip is not 
compatible with what is usually observed from an HMXB in eclipse. In order to gain further insights on the nature of the dip, we thus also looked at the pulsation 
characteristics of this source by using all publicly available IBIS/ISGRI and JEM-X data from the \emph{INTEGRAL} satellite. 

\section{Observation and Analysis}
\subsection{\emph{Suzaku}}
\label{sect:suzaku data analysis}
\emph{Suzaku} \citep{M07} is a broad-band X-ray observatory which covers
the energy range of 0.2-600 \rmfamily{keV}. It has two main instruments: the X-ray Imaging Spectrometer {XIS} \citep{K07}, covering the 0.2-12
\rmfamily{keV} energy range, and the Hard X-ray Detector (HXD). The latter comprises PIN diodes \citep{T07}, covering 
the 10-70 \rmfamily{keV} energy range, and GSO crystal scintillator
detectors, covering the 70-600 \rmfamily{keV} energy range. 
The XIS consists of four CCD detectors of which three are front illuminated 
(FI) and one is back illuminated (BI). Only three out of the four XIS units (XIS 0,1 and 3) are currently 
operational. \\
4U 0114+65 was observed with \emph{Suzaku} on 2011 July 21-22 (OBSID `406017010'). The observation was carried out at the `XIS nominal'
pointing position and the effective exposure time of the XIS and PIN was of 106.5 and 88.5 \rm{ks}, respectively. The XIS units were operated
in `standard' data mode with the `Window 1/4' option (providing a time resolution of 2 \rm{s}). \\
For the XIS and HXD data, we used the filtered cleaned event files which are obtained using the pre-determined screening criteria as suggested in 
the Suzaku ABC guide \footnote{http://heasarc.gsfc.nasa.gov/docs/suzaku/analysis/abc/}. XIS light curves and spectra were extracted from the XIS data by
choosing circular regions of $3^{'}$ radius from the source position.
Background light curves and spectra for the XIS were extracted by selecting regions of
the same size in a portion of the CCD that did not contain any source photons.
For the HXD/PIN, simulated `tuned' non
 X-ray background event files (NXB) corresponding to the month and year of the respective observations 
were used to estimate the non X-ray background \footnote{http://heasarc.nasa.gov/docs/suzaku/analysis/pinbgd.html}\citep{F09}. \\
The timing analysis of the \emph{Suzaku} data was performed on the XIS and PIN light curves after applying
barycentric corrections to the event data files (we used the \texttt{FTOOLS} task
`aebarycen'). These files were also corrected for dead time effects by using the \texttt{FTOOLS} task `hxddtcor'.
Light curves were extracted from the XIS data with the highest available time resolution of 2 \rm{s}. 
We summed 
the background subtracted XIS 0, 1 and 3 light curves and obtained a single background corrected light curve for all the XIS units. 
The PIN light curves were extracted with a resolution of  1 \rmfamily{s}
and the corresponding background was evaluated by generating similar light curves from the simulated background files.\\
The XIS and PIN light curves comprise 14 neutron star rotations, as shown in Figure \ref{lcurve}. The average source count-rate in the two instruments was 2.69 c/s and 0.85 {c/s}, respectively.\\
The XIS spectra were extracted with 2048 channels and the PIN spectra with 255 channels. 
Response files for the XIS were created using CALDB version `20140701'. For the HXD/PIN spectrum, response files corresponding to the epoch 
of the observation were obtained from the \emph{Suzaku} guest observer
facility\footnote{http://heasarc.nasa.gov/docs/heasarc/caldb/suzaku/}. 
\subsubsection{Timing Analysis}
\label{sect:suzaku timing analysis}
\begin{figure*}
\begin{center}
\includegraphics[width=9cm,height=12cm,angle=-90]{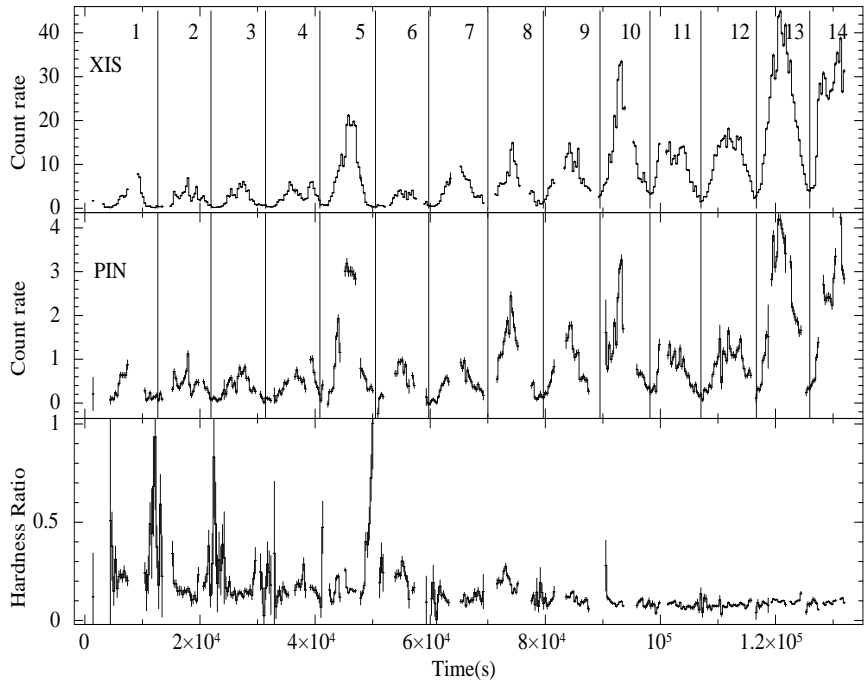}
\end{center}
\caption{\rmfamily{Background subtracted \emph{Suzaku} light curves of 4U 0114+65. The time bin is 300 \rm{s}. 
The upper and middle panels show the XIS and PIN data, respectively. The hardness ratio of the XIS and PIN data is reported in the bottom panel. The start 
time of the light curves is MJD 55763.4805}}
\label{lcurve} 
\end{figure*}

The source PIN and XIS light curves are reported in Figure \ref{lcurve} (the binsize is 300 \rm{s}).
The upper and middle panel of the figure show the XIS and PIN data respectively. 
The lower panel show the hardness ratio between the PIN and XIS. Pulsations can be clearly seen even during the lowest 
X-ray emission time interval, i.e., the dip.
The average count rates measured during dip and and out of the dip differ only by a
factor of $\sim$ 2-4.

\begin{figure*}
\includegraphics[width=9cm,height=11cm,angle=0]{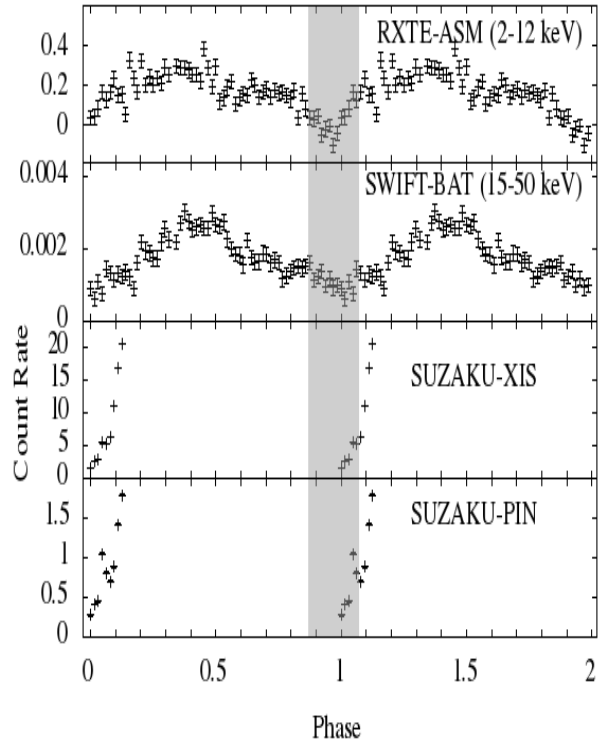}
\caption{\rmfamily{Light curves of \emph{RXTE}-ASM, \emph{Swift}-BAT, XIS and HXD-PIN folded
 at the orbital period of 4U 0114+65 (the initial epoch is chosen to be MJD 55763.4805). A complete eclipse is detected only in the ASM data, which covers a softer X-ray
 energy range than BAT. (Note that the negative count rate in the ASM data is only apparent which arises from the automated background subtraction of the 
 ASM lightcurve and correspond to the time when the source was too faint to be detected by it). The shaded region shows the ingress and egress time of the dip (as seen in the ASM data) lasting for nearly one day each. }}
\label{bat_xis_pin} 
\end{figure*}

\begin{figure*}
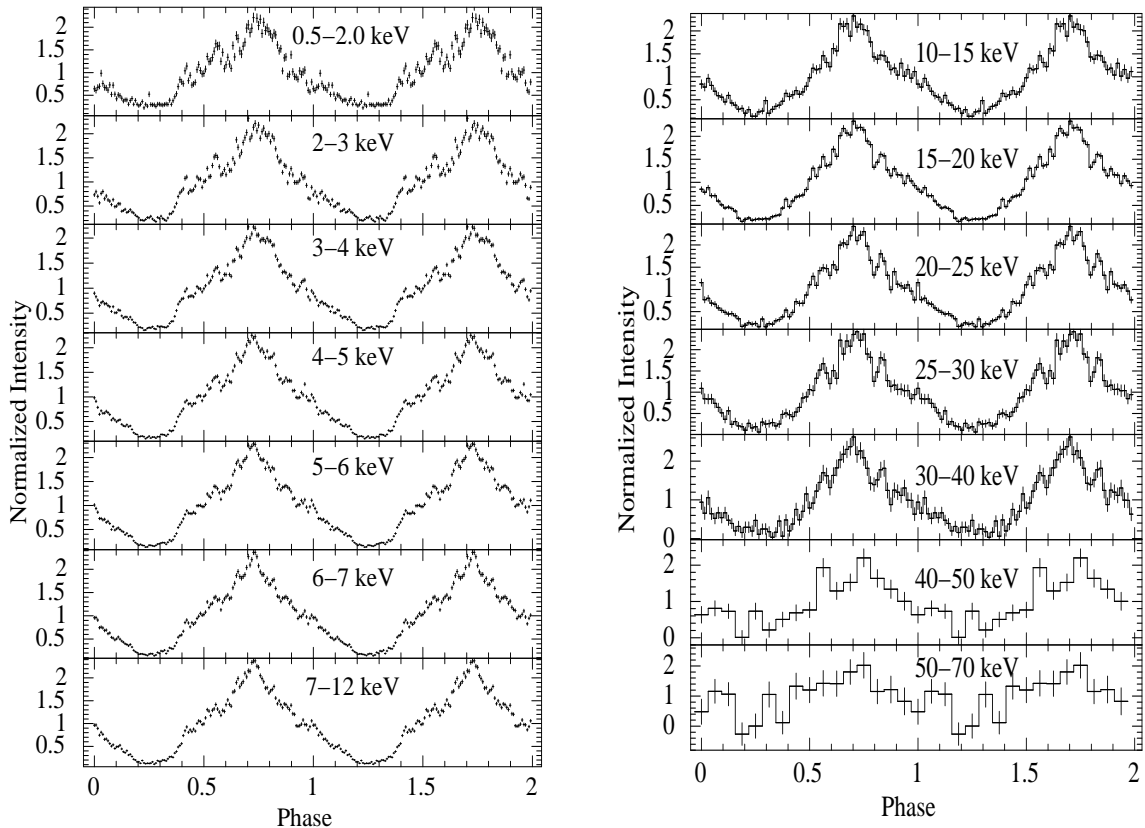

\includegraphics[width=11cm,height=8cm,angle=-90]{pp_er_xis1.ps}
\includegraphics[width=10.9cm,height=7.9cm,angle=-90]{pp_er_pin1.ps}
\caption{\rmfamily{The left and right panels show the energy resolved pulse profiles of the XIS and PIN respectively. We used a spin period of 9391.19 \rm{s} and an initial 
epoch TJD (Truncated Julian Date) 15763. The pulse profiles do not show any obvious energy dependence.}}
\label{pp_er}
\end{figure*}

To determine the orbital phase of the \emph{Suzaku} observation, we folded the \emph{Suzaku}-XIS and PIN light curves together with the 
\emph{Swift}-BAT\footnote{http://swift.gsfc.nasa.gov/results/transients/weak/3A0114p650/} and
\emph{RXTE}-ASM\footnote{ftp://legacy.gsfc.nasa.gov/xte/data/archive/ASMProducts/definitive\_1dwell/lightcurves/}
long term light curve of 4U 0114+65 at an orbital period of 11.596 days. The latter was obtained from the \emph{RXTE}-ASM light curve using the \texttt{FTOOLS} task `efsearch'. The folded 
light curves  in Figure \ref{bat_xis_pin} clearly show that the \emph{Suzaku} observation was carried out during the X-ray dip.
We also extracted the energy resolved pulse profiles of the source, normalized at the average source intensity for the XIS and PIN data. 
As shown in Figure \ref{pp_er}, the profiles do not show any significant energy dependence and pulsations are always detected up to 70 \rm{keV}.   

\subsubsection{Spectral Analysis}
\label{sect:spectral analysis}
\begin{figure*}
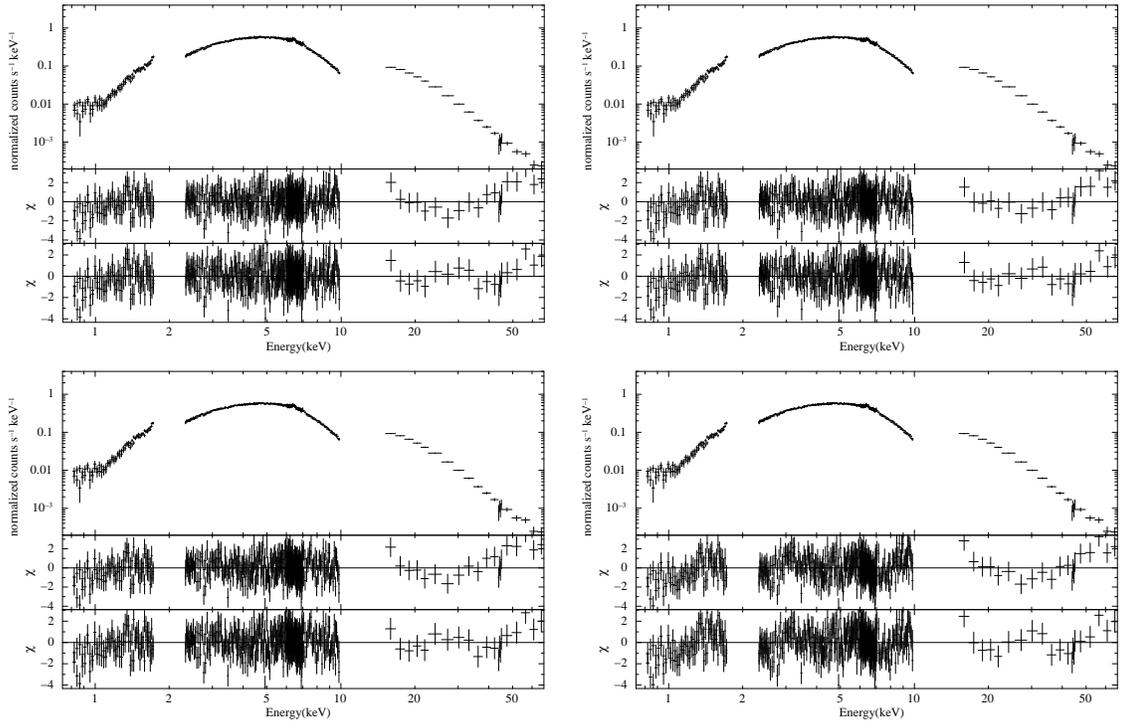

\centering
\includegraphics[scale=0.3,angle=-90]{high_final.ps}
\includegraphics[scale=0.3,angle=-90]{npex_final.ps}
\includegraphics[scale=0.3,angle=-90]{final_cutoff.ps}
\includegraphics[scale=0.3,angle=-90]{final_comptt.ps}
\caption{\rmfamily{{Time averaged X-ray spectra of 4U 0114+65 for \emph{Suzaku}-XIS and PIN data of 4U 0114+65 fitted with the four models measured in the text. 
The second panel in each figure shows the residuals from the fits when the cyclotron absorption component is not included. The
third panel respresents the same residuals after the inclusion of the absorption feature.}}}
\label{spectra_cyc}
\end{figure*}

We performed a time averaged spectral analysis of the X-ray emission from 4U 0114+65 using the two XIS units (XIS 0 and 3) and the PIN data. 
Data from the back illuminated XIS 1, were not included in the spectral analysis as the cross-normalizations obtained between the XIS 0 and the XIS 1 
were larger than those suggested in the \emph{Suzaku} ABC guide.
 Additionally, we also noticed systematic differences in the spectral fitting of the BI data
compared to the other XIS units. 
Such differences between the BI and FI XIS units have been previously reported, e.g., in the case of GX 301-2 \citep{suchy2011} and 
IGR J16318-4848 \citep{barra2009}. 
In these situations, it is more convenient to use only  the data from the FI XIS units, given the greater sensitivity of the the latter chips 
above 2 \rm{keV} compared to the BI XIS unit.\\
Spectral fitting was performed by using \texttt{XSPEC} v12.8.2.
For the spectral analysis we have chosen the energy range 0.8-10 \rmfamily{keV}
for the XIS units and 15.0-70.0 \rmfamily{keV} for the PIN, respectively.
Artificial residuals are known to arise in the XIS spectra around the Si
edge and Au edge. The energy range 1.75-2.23 \rmfamily{keV} is thus usually discarded for the spectral analysis.
We fitted all 
spectra simultaneously with all parameters tied, except the relative 
instrument normalizations which were kept free to vary. The 2048 channel XIS spectra were rebinned by a factor of 6 up to 5 \rmfamily{keV},
by a factor of 2 from 5-7 \rmfamily{keV}, 
and by a factor of 14 in the remaining energy range.
The PIN spectra were binned by a factor of 4 at energies $\leq$ 22 \rmfamily{keV}, by 8 in the energy range 22-45 \rmfamily{keV}, and by 12 in the remaining 
part of the instrument passband. 
To fit the spectral continuum, we used the continuum models commonly used for HMXBs, such as a
power-law model modified with an exponential cutoff (CUTOFFPL), a power law with a high energy cutoff (HIGHECUT; \citealt{W83, CO01}), 
a negative and positive power law with exponential continuum model (NPEX; \citealt[]{M95, makishima1999}) and a physical Comptonization model (COMPTT; \citealt{TI94}). In addition to the interstellar absorption along our line of sight, we also used a partial covering component to account for local absorption. 
A soft excess below 2 \rm{keV} remained visible in the residuals from the fits with all above models except COMPTT. 
Where significantly detected, the soft excess was accounted for with the addition of a blackbody component. Two gaussian components were 
added to fit the iron K$_\alpha$ and K$_\beta$  emission lines. 
An evident feature is 
seen in the residuals from all the fits between 20-30 \rm{keV}, irrespectively of the model used for the continuum. 
The inclusion of an absorption cyclotron feature (CYCLABS) with a width fixed at 9.8 \rm {keV} \citep{bonning2005} slightly improved the fit. 
To test further, the significance of the cyclotron line detected, we used the F-test routine available in the IDL package 
mpftest\footnote{http://cow.physics.wisc.edu/$\sim$craigm/idl/down/mpftest.pro} \citep{D13}. The probability of chance improvement (PCI) is evaluated for the HIGHECUT model used to fit the spectrum with 
and without the cyclotron line. The estimated PCI value after addition of cyclotron line component to the HIGHECUT is $\sim$ 38.4 \% . 
The best-fit spectral parameters for 4U 0114+65 for all considered models are summarized in Table \ref{spec_par}.
The time-averaged spectra for different models and their residuals with and without the inclusion of the CYCLABS component are shown in Figure \ref{spectra_cyc}. At energies above 50 \rm{keV}, 
the residuals also suggested the presence of a high energy tail in 4U 0114+65 \citep[]{wei2011,denhartog2006}, the detailed fitting of which we have not 
been able to carry out since the HXD/PIN spectrum is limited to 70 \rm{keV}. \\
\begin{table*} 
\scriptsize
\caption{\rm{Best-fit parameters of the time averaged spectra of 4U 0114+65 for different models extracted from the $Suzaku$ observation.
 All reported uncertainties are at 90 per cent confidence level.}}
\label{table}
\begin{center}
\begin{tabular}{|c | c | c |c |c |c |c |c |}
\hline
Parameter & HIGHECUT & NPEX  & CUTOFFPL & COMPTT  \\
\hline
${N_{\rm H1}^{_a}}$  & $4.81_{-0.22}^{+0.16}$ & $4.78_{-0.20}^{+0.16}$ & $4.92_{-0.21}^{+0.17}$ & $1.83_{-0.04}^{+0.05}$ \\ \\
${N_{\rm H2}^{_a}}$  & $23.54_{-10.23}^{+16.18}$ & $24.08_{-4.51}^{+4.45}$ & $22.51_{-4.01}^{+3.85}$ & $157.15_{-23.44}^{+21.98}$ \\ \\
CvrFract & $0.12_{-0.04}^{+0.05}$ & $0.25_{-0.03}^{+0.03}$ & $0.27_{-0.03}^{+0.03}$ & $0.41_{-0.07}^{+0.06}$ \\ \\
PhoIndex & $0.77_{-0.06}^{+0.05}$ & $0.58$*$_{-0.09}^{+0.10}$ & $0.74_{-0.06}^{+0.05}$ & -- \\ \\
Powerlaw(norm)$^{_b}$  & $0.011_{-0.001}^{+0.001}$ & $0.013_{-0.001}^{+0.001}$ & $0.015_{-0.001}^{+0.001}$ & --\\ \\
Ecut (keV) & $5.94_{-0.26}^{+0.23}$  & $12.26_{-1.69}^{+2.49}$ & $18.06_{-1.78}^{+2.20}$ & -- \\ \\
Efold (keV) & $19.21_{-1.87}^{+2.17}$ & -- & -- & --\\ \\
compTT ($T_{0}$) (keV) & -- & -- & -- & $1.61_{-0.04}^{+0.04}$\\ \\
compTT(kT) (keV) & -- & -- & -- & $10.65_{-1.13}^{+2.15}$\\ \\
compTT(taup) & -- & -- & -- & $3.06_{-0.53}^{+0.39}$\\ \\
compTT(norm) & -- & -- & -- & $0.0056_{-0.0008}^{+0.0008}$\\ \\
bbody(kT) (keV) & $0.137_{-0.005}^{+0.007}$ & $0.138_{-0.005}^{+0.006}$ & $0.134_{-0.005}^{+0.006}$ & --  \\ \\
bbody(norm) $^{_c}$  & $0.018_{-0.007}^{+0.009}$ & $0.017_{-0.006}^{+0.009}$ & $0.024_{-0.009}^{+0.013}$ & --\\ \\
Cyclabs$^{**}$ (keV) & $28.17_{-3.36}^{+2.62}$ & $30.34_{-5.79}^{+3.41}$ & $26.64_{-3.95}^{+2.95}$ & $28.74_{-2.93}^{+3.14}$ \\ \\
Depth(keV)& $0.19_{-0.08}^{+0.08}$ & $0.16_{-0.09}^{+0.11}$ & $0.17_{-0.07}^{+0.08}$ & $0.27_{-0.08}^{+0.09}$\\ \\
K$_\alpha$ line (keV) & $6.42_{-0.01}^{+0.01}$ & $6.42_{-0.01}^{+0.01}$ & $6.42_{-0.01}^{+0.01}$ & $6.42_{-0.01}^{+0.01}$\\ \\
Equivalent Width for K$_\alpha$ line (keV) & $0.029_{-0.004}^{+0.004}$ & $0.029_{-0.003}^{+0.003}$ & $0.029_{-0.003}^{+0.003}$ &  $0.029_{-0.003}^{+0.003}$\\ \\
K$_\beta$ line (keV) & $7.09_{-0.04}^{+0.03}$ & $7.12_{-0.04}^{+0.03}$ & $7.13_{-0.04}^{+003}$ & $7.08_{-0.01}^{+0.01}$\\ \\
Equivalent Width for K$_\beta$ line (keV) & $0.012_{-0.004}^{+0.004}$ & $0.012_{-0.003}^{+0.003}$ & $0.014_{-0.004}^{+0.004}$ &  $0.0029_{-0.0028}^{+0.004}$\\ \\
Flux${_{XIS}^{d}}$ & $1.57_{-0.04}^{+0.04}$  & $1.56_{-0.02}^{+0.02}$ & $1.51_{-0.04}^{+0.04}$ & $1.52_{-0.04}^{+0.04}$ \\ \\
Flux${_{PIN}^{d}}$ & $3.95_{-0.08}^{+0.08}$ & $3.97_{-0.02}^{+0.02}$ & $3.91_{-0.07}^{+0.07}$ & $3.95_{-0.08}^{+0.08}$  \\ \\
$\chi^{2}_{\nu}$/d.o.f (with CYCLABS) & 1.24/462 & 1.27/462 & 1.28/463 & 1.48/467 \\ \\
 $\chi^{2}_{\nu}$/d.o.f (without CYCLABS) & 1.28/464 & 1.29/464 & 1.32/465 & 1.55/469\\ \\
 \hline
$^{_a}$In units of $10^{22}$ atoms $cm^{-2}$ \\
$*$ Photon-index of the second power-law component of the NPEX is frozen to 2.0 \\
$^{**}$Width frozen at 9.8 \rm{keV} \citep{bonning2005} \\
$^{_b}$ In units of photons $keV^{-1}$ $cm^{-2}$ $s^{-1}$ at 1 \rm {keV} \\
$^{_c}$ In units of ${L_{39}}$/${D_{10}}$ \\
$^{_d}$ In units of $10^{-10}$ $ergs$ $cm^{-2}$ $s^{-1}$

\end{tabular}
\end{center}
\label{spec_par}
\end{table*}
To perform a time resolved spectral analysis of the emission within the dip, we divided the
light curve into fourteen segments. Each segment is defined to be the time elapsed between two successive minima in the light curve, as is depicted in Figure \ref{lcurve}. This way, 
the average time interval between the segments turn out to be $\sim$ 9355 \rm{s},
which is nearly equal to the spin period of the pulsar ($\sim$ 9391 s).

\begin{figure*}
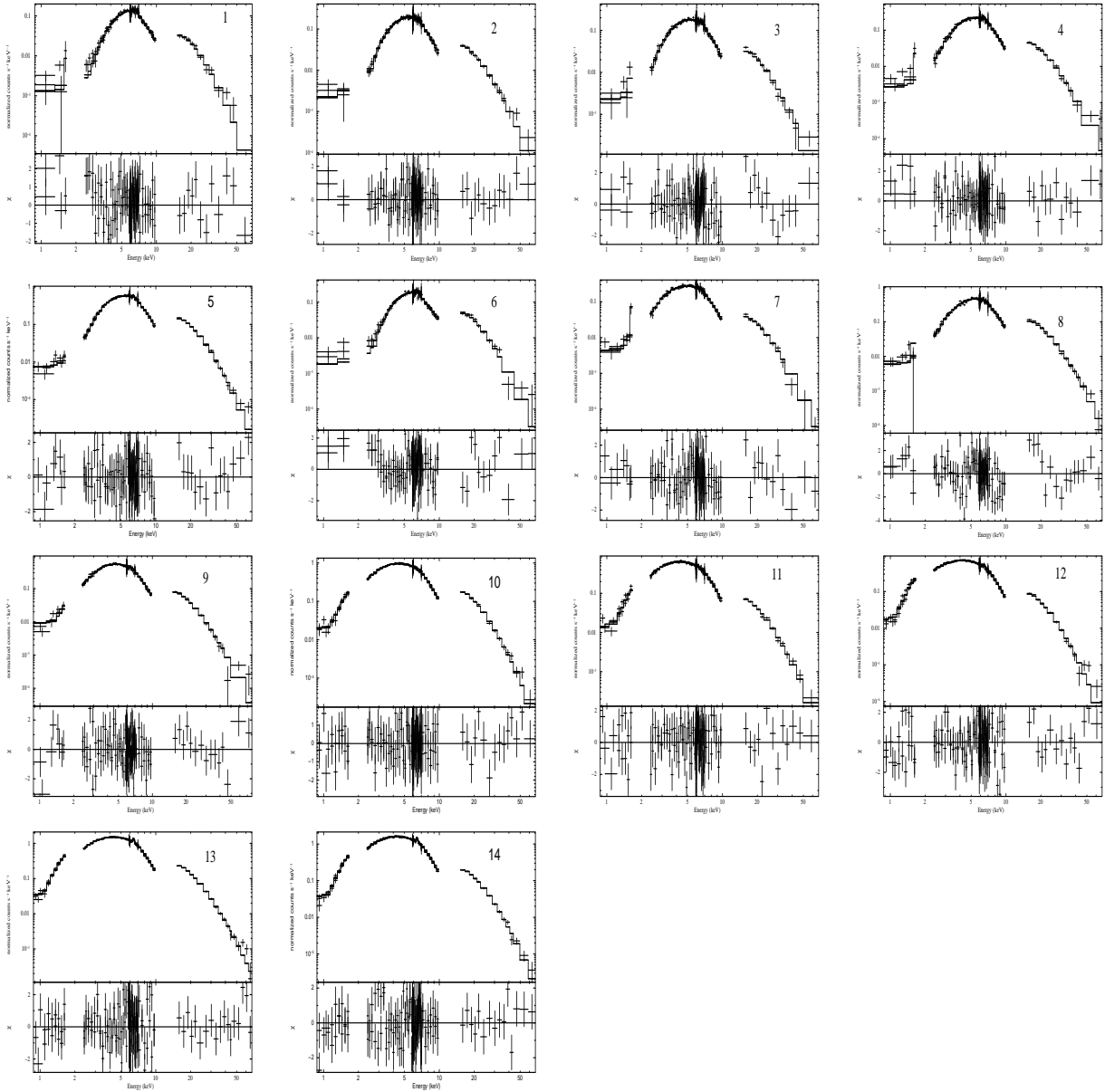

\begin{center}$
 \begin{array}{ccccccc}
\includegraphics[height=1.5in, width=1.5in, angle=-90]{1.ps} &
\includegraphics[height=1.5in, width=1.5in, angle=-90]{2.ps} &
\includegraphics[height=1.5in, width=1.5in, angle=-90]{3.ps} &
\includegraphics[height=1.5in, width=1.5in, angle=-90]{4.ps} \\
\includegraphics[height=1.5in, width=1.5in, angle=-90]{5.ps} &
\includegraphics[height=1.5in, width=1.5in, angle=-90]{6.ps} &
\includegraphics[height=1.5in, width=1.5in, angle=-90]{7.ps}&
\includegraphics[height=1.5in, width=1.5in, angle=-90]{8.ps} \\
\includegraphics[height=1.5in, width=1.5in, angle=-90]{9.ps}&
\includegraphics[height=1.5in, width=1.5in, angle=-90]{10.ps} &
\includegraphics[height=1.5in, width=1.5in, angle=-90]{11.ps}&
\includegraphics[height=1.5in, width=1.5in, angle=-90]{12.ps} \\
\includegraphics[height=1.5in, width=1.5in, angle=-90]{13.ps}&
\includegraphics[height=1.5in, width=1.5in, angle=-90]{14.ps} &
\end{array}$
\end{center}
\caption{\rm{Time resolved spectra of  4U 0114 +65 fit with the HIGHECUT model. Residuals from the fits are shown in the bottom panel of each figure}}
\label{tr_spectra}
\end{figure*}

\begin{figure*}
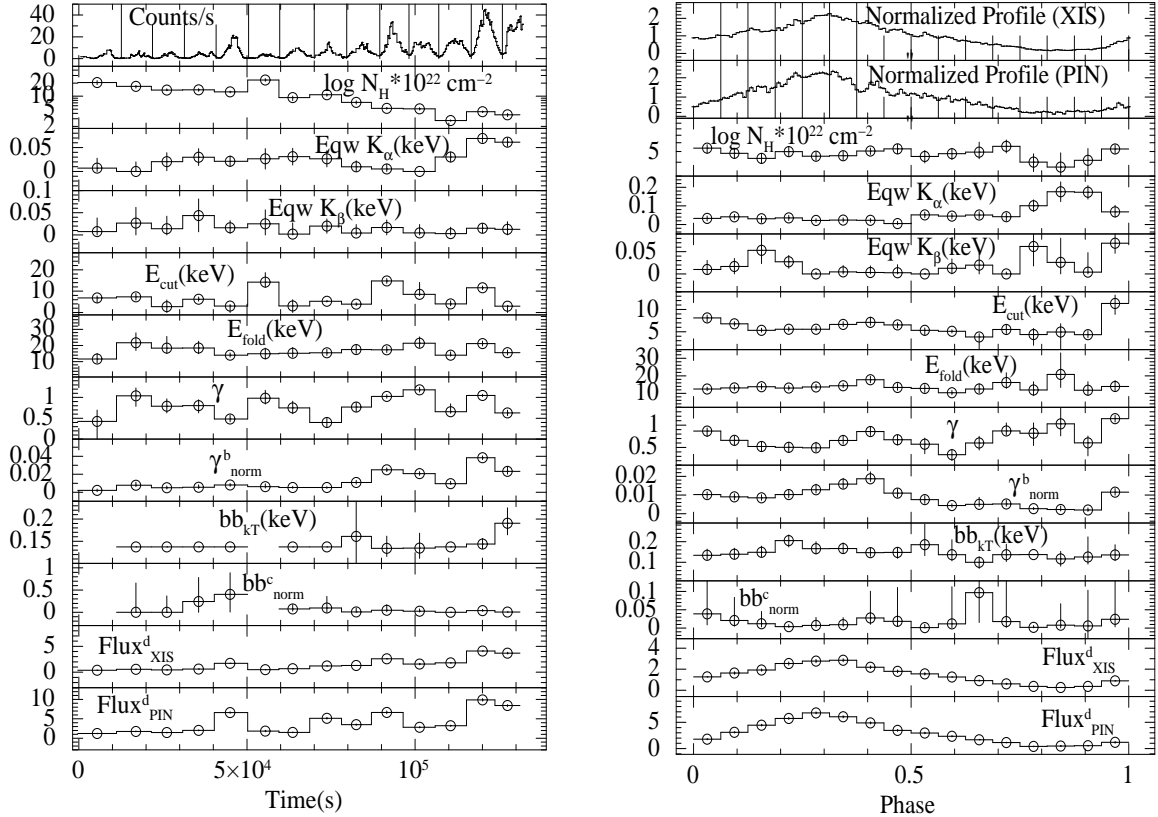

\begin{flushleft}
\includegraphics[width=11cm,height=8cm,angle=-90]{time_lc_pcfabs_ref1.ps}
\includegraphics[width=11cm,height=8cm,angle=-90]{phase_pp_param_ref1.ps}
\end{flushleft}
\caption{\rmfamily{Left: Variations of the source spectral parameters (for the HIGHECUT model) with time 
(t=0 corresponds to MJD 55763.4805). Right: Variations of the source spectral parameters with pulse phase (for the HIGHECUT model). 
$^{_b}$ = In units of photons $keV^{-1}$ $cm^{-2}$ $s^{-1}$ at 1 \rm {keV}, $^{_c}$ = In units of ${L_{39}}$/${D_{10}}$; $^{d}$ = In units of $10^{-10}$ 
$ergs$ $cm^{-2}$ $s^{-1}$}}
\label{param} 
\end{figure*}

The time resolved spectra (shown in Figure \ref{tr_spectra}) could be well fit with the HIGHECUT model, which also 
provided the best description of the time averaged spectrum. 
The variations of the spectral parameters with time 
obtained for the HIGHECUT model are shown on the left side of Figure \ref{param}. 
\\
For the spin phase resolved spectral analysis, we extracted the source spectra in 16 phase bins. Each phase resolved spectrum was again fitted with the HIGHECUT model. 
The variation of the spectral parameters with the spin phase is shown on the right hand side of Figure \ref{param}.

\begin{figure*}
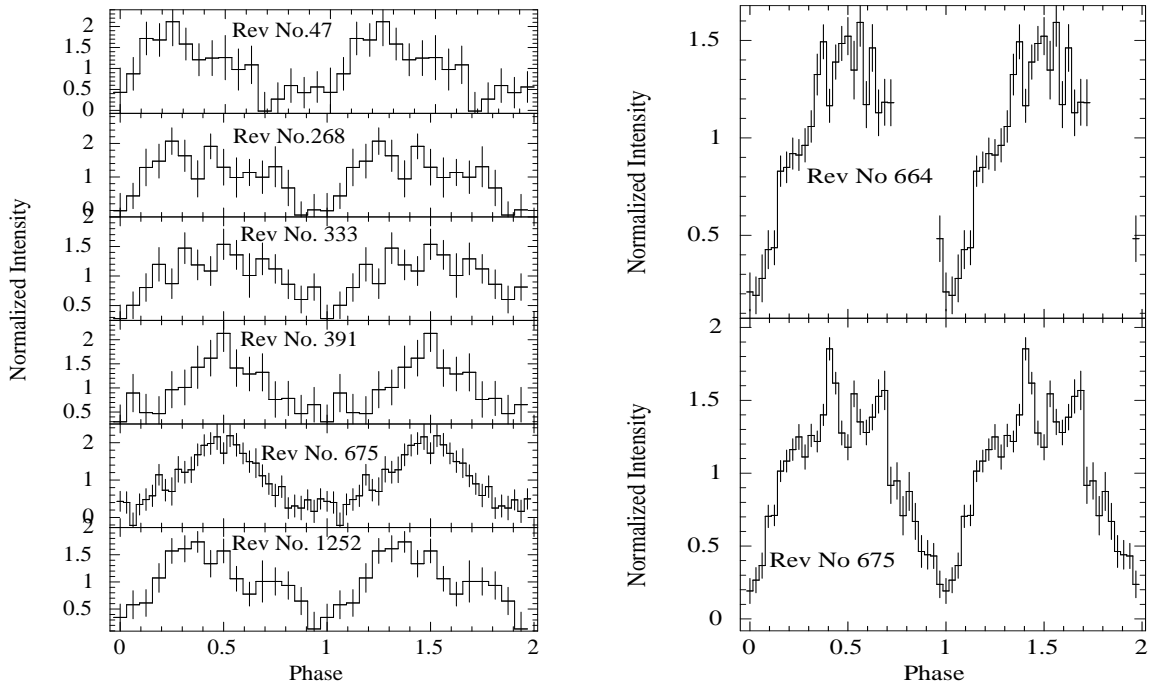

\begin{center}
\includegraphics[width=9cm,height=8cm,angle=-90]{integral_pp.ps}
\includegraphics[width=9cm,height=8cm,angle=-90]{pp_jx1_3-35.ps}
\end{center}
\caption{\rmfamily{Left: Pulse profiles extracted from the IBIS/ISGRI data (20-40 \rm{keV}). Right: Pulse profile extracted from the JEM-X data
(3-35 \rm{keV}). In both panels, the epoch chosen for the folding is arbitrary.
}}
\label{integral_pp} 
\end{figure*}
From the time resolved spectral analysis of the source, we obtained the following results:
\begin{enumerate}
\item The partial covering absorption component was not needed to fit the time resolved spectra. The measured variation of N$_{\rm H}$ with time 
suggests the presence of clumps in the neutron star surroundings as in the case of GX 301-2 \citep{mukherjee2004}, Vela X-1 \citep{furst2010}, Cen X-3 \citep{naik2011} and OAO 1657-415 \citep{pradhan2014}. 
We found that the N$_{\rm H}$ was higher at the beginning of the observation  (2 $\times$ $10^{23}$ atoms $cm^{-2}$) and decreased by a factor of 10 
towards the end of the \emph{Suzaku} observation window.
\item For low flux segments like (2-5 and 7-8 and 12), the blackbody temperature had to be frozen to the time-averaged value while allowing the blackbody normalization 
to vary. For low flux segments 1 and 6, we could not constrain the black body normalization even by freezing the blackbody temperature to time-averaged value.
\item Residuals from the fits around the cyclotron line energy are not detected in most of the time resolved segments (due to low statistics) except for the segments 3, 5 
and 8. In these cases, however, the statistics was far too low to perform a meaningful test about the statistical significance of the possible cyclotron line.
\end{enumerate}
From the results of the phase resolved spectral analysis we conclude that:
\begin{enumerate}
\item As for the time resolved spectral analysis, the partial covering absorption component was not required to fit the for phase resolved spectra. 
\item Only the photon index and the normalization of the power-law component show some moderately significant variability with the pulse phase.
\item Residuals around the energy of the cyclotron line could not be detected, most likely due to the limited statistics compared
to the phase averaged spectrum.
\end{enumerate} 
Finally, we showed through our timing analysis of the \emph{Suzaku} data that pulsations are detected in the X-ray emission of the source also during the time 
interval corresponding to the dip.
\subsection{\emph{INTEGRAL}/IBIS and JEM-X light curves}

We used all publicly available INTEGRAL data collected in the direction of 4U\,0114+65 since the earliest science operations
of the mission. This data-set comprises `science windows' (SCWs), i.e. pointings with typical durations of $\sim$2-3~ks, carried
out from the satellite revolution 25 to 1471. We analyzed all data from the IBIS/ISGRI \citep{lebrun03} and the two JEM-X \citep{lund03} instruments
by using version 10.1 of the Off-line Scientific Analysis software (OSA) distributed by the ISDC \citep{courvoisier03}.
To limit the IBIS/ISGRI calibration uncertainties, we selected only SCWs during which
the source was located to within 12~deg from the center of the IBIS FoV \citep{ubertini03}. 
The total effective exposure time was of 8.9 Ms for IBIS/ISGRI, 1.4 Ms for JEM-X1, and 520 ks for JEM-X2.
From all the available \emph{INTEGRAL} IBIS/ISGRI and JEM-X observations of 4U 0114+65, we extracted the source light curves with a time binning of 100 \rm{s} in the 
energy range 20-40 \rm{keV} and 3-35 \rm{keV}, respectively.
We then searched for pulsations during those observations that were carried out at orbital phases corresponding to the X-ray dip ($\pm$ 1 day). 
Except for those cases where the statistics of the data was relatively low, pulsations were seen in most of the observations and a few examples of pulse profiles are shown in Figure \ref{integral_pp}.
\section{Discussion}
In this paper, we used \emph{Suzaku} and \emph{INTEGRAL} data to study the X-ray emission from 4U 0114+65 during the orbital phase corresponding to its periodic
X-ray dip. Such dip has been interpreted in the past as an eclipse of the neutron star hosted in this system by its supergiant companion. However, our analysis revealed
some anomalous properties of the source X-ray emission that would not support such interpretation. In particular we noted that: 
\begin{enumerate}
\item The orbital intensity profile of 
4U 0114+65 in the proximity of the X-ray dip is different from that of other eclipsing HMXBs which typically show sharp eclipse ingresses and egresses 
in hard X-rays and a more gradual variation of their intensity in soft X-rays owing to absorption in the stellar wind. 
On the contrary, we showed that 4U 0114+65 exhibits an eclipse-like profile at the lower energies (\emph{RXTE}-ASM) with the ingress and egress time of 
the dip being nearly a day-long each, and a more gradual variation in hard X-rays (\emph{Swift}-BAT). 
\item The X-ray spectra of HMXBs during eclipses are characterized by the presence of prominent iron lines with large equivalent widths, (see e.g., the cases of 
 Vela X-1: \citealt[]{nagase94}; Cen X-3: \citealt[]{nagase92, naik2011}). The spectrum of 4U 0114+65 does not show such feature. 
 \item The spectral index during eclipse should be higher than out of eclipse phases, since during eclipse, we will be observing only the 
 scattered emission in the wind. As an example, in the \emph{Suzaku} observation of Cen X-3 when the pulsar was observed during eclipse and out of eclipse
  phases, the photonindex correlates well with those orbital phases (Figure 6 of \citealt{naik2011}). We see no such correlation here. 
\item Eclipse ingresses and egresses of HMXBs are often characterized by extremely large increases/decreases of the absorption column densities due to the presence of 
the dense wind of the supergiant 
star along the line of sight to the observer (N$_{H}$ $\geq$ $10^{24}$ atoms $cm^{-2}$). A soft excess should not be observable in these cases, as the X-ray emission at 
energies $\leq$ 2 \rm{keV} is strongly extinguished.
In the case of 4U 0114 +65, we observe only a moderate increase of the absorption column density in the dip compared to other orbital phases and also the soft 
excess below $\sim$ 2 \rm {keV} remains clearly detectable in the averaged \emph{Suzaku} spectrum and in those segments where the flux is not too low. 
\item Pulsations were clearly detected during all available \emph{INTEGRAL} observations carried out during the X-ray dip and the \emph{Suzaku} observations, 
even though in these cases the source was supposed to be in eclipse. 
\end{enumerate} 

Different works in the past have investigated the origin of the X-ray dip in 4U 0114+65. \cite{hall2000} and \cite{corbet1999} reported a definite truncation of X-ray signal 
near orbital phase 0 indicative of an eclipse which is further supported by an increase in column density near the same phase \citep{hall2000}. 
On the other hand, neither \cite{crampton1985} nor \cite{farrell2008} found convincing evidence for the presence of an X-ray eclipse in the \emph{RXTE}/PCA and
optical data. \cite{grundstrom2007} was the first to note that the minimum in the X-rays
occurs very close to the predicted phase of the supergiant inferior conjunction. 
In agreement with their findings, we thus propose that, as the system is characterized by an eccentric orbit, the accretion rate onto the neutron star
is modulated along the orbital phase, leading to variations in both soft and hard X-rays. It is not an eclipsing
binary, but at the inferior conjunction of the companion star, increased absorption in the stellar wind causes a dip in the soft X-ray orbital light curve,
which is also seen in the form of a larger N$_H$ in the beginning of the \emph{Suzaku} observation. It shall be remarked that the detection of X-ray pulsations 
during the dip cannot be used to distinguish between the idea proposed above and the eclipse model. Indeed, the long spin period of the pulsar hosted in 4U 0114+65 would 
allow us to detect pulsations even when the source is eclipsed, as these would not be washed out in the emission reprocessed by the stellar wind. \\
Finally we suggest that the variations of the absorption column density (see Figure \ref{param}) recorded during the \emph{Suzaku} observation on time-scales of few 
thousand seconds might be due to the presence of clumps in the wind of the supergiant companion as observed also in other similar systems (see, e.g., GX 301-2, Vela X-1,
 Cen X-3 and OAO 1657-415; \citealt{mukherjee2004, furst2010, naik2011, pradhan2014, walter2015} ) 

\section{Acknowledgement}
This research has made use of data obtained from the \emph{Suzaku} satellite, a collaborative mission between the space agencies of Japan (JAXA) and the USA (NASA). 
The data for this work has been obtained through the High Energy Astrophysics Science Archive (HEASARC) Online Service provided by NASA/GSFC. 
We have also made use of public light curves from \emph{Swift} and \emph{RXTE} site. PP is thankful to UGC for their financial support under `Minor Research Project' 
and also to the hospitality provided by Raman Research Institute, Bangalore for carrying out this work. TMB acknowledges support from INAF-PRIN 2012-6.

\bibliography{4u0114_ref5.bib}{}
\bibliographystyle{mn2e}

\end{document}